\documentclass[10pt,conference]{IEEEtran}

\usepackage{graphicx}
\usepackage[many]{tcolorbox}

\usepackage[noadjust]{cite}
\usepackage{amsmath,amssymb,amsfonts}
\usepackage{textcomp}
\usepackage{xcolor,soul}
\usepackage[normalem]{ulem}
\usepackage{amsmath}
\usepackage{mathtools,amssymb,latexsym,amsfonts,stmaryrd}
\usepackage{pbox}
\usepackage{xfrac}
\usepackage{booktabs}
\usepackage{xcolor}
\usepackage{threeparttable}
\usepackage{amsthm}
\usepackage{multirow}
\usepackage{tikz}
\usepackage{mathrsfs}
\usepackage{mathpartir}
\usepackage{algorithm}
\usepackage{url}
\usepackage{syntax}
\usepackage{framed}
\usepackage[noend]{algpseudocode}
\usepackage{flushend}
\usepackage{listings}
\usepackage{moresize}
\usepackage{wrapfig}
\usepackage{xcolor}
\usepackage{colortbl}
 
\usepackage{seqsplit}
\usepackage{alltt}
\usepackage{xspace}
\usepackage{makecell}
\usepackage{subcaption}
\usepackage{pifont}

\definecolor{pinegreen}{HTML}{01796f}

\newcommand{\leaked}{\includegraphics[scale=0.075]{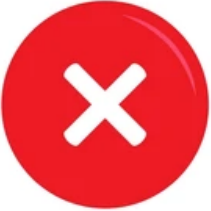}}
\newcommand{\secure}{\includegraphics[scale=0.075]{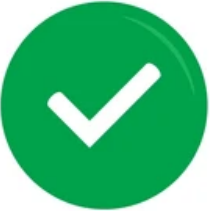}}

\DeclareMathAlphabet{\mathcal}{OMS}{cmsy}{m}{n}

\usepackage{amsmath, listings, amsthm, amssymb, proof, xspace}

\usepackage{thmtools}

\declaretheoremstyle[spaceabove=\topsep,notefont=\normalfont\itshape]{mystyle}

\newcommand{\revise}[2]{{\color{red}{\ifx&#1&\else- #1\fi}}{\color{blue}{\ifx&#2&\else#2\fi}}}%

\usetikzlibrary{arrows,patterns, decorations.pathreplacing}

\usepackage[colorlinks]{hyperref}   
\AtBeginDocument{%
  \hypersetup{pdfborder={0 0 1}, urlcolor=.}  
}

\definecolor[named]{ACMBlue}{cmyk}{1,0.1,0,0.1}
\definecolor[named]{ACMYellow}{cmyk}{0,0.16,1,0}
\definecolor[named]{ACMOrange}{cmyk}{0,0.42,1,0.01}
\definecolor[named]{ACMRed}{cmyk}{0,0.90,0.86,0}
\definecolor[named]{ACMLightBlue}{cmyk}{0.49,0.01,0,0}
\definecolor[named]{ACMGreen}{cmyk}{0.20,0,1,0.19}
\definecolor[named]{ACMPurple}{cmyk}{0.55,1,0,0.15}
\definecolor[named]{ACMDarkBlue}{cmyk}{1,0.58,0,0.21}
\hypersetup{
  colorlinks,
  linkcolor=ACMPurple,
  citecolor=ACMPurple,
  urlcolor=ACMDarkBlue,
  filecolor=ACMDarkBlue
}

\usepackage{inconsolata} 
\usepackage{biolinum} 

\usepackage{microtype}

\usepackage[noadjust]{cite}

\newcommand{\F}{Fig.}
\newcommand{\E}{Eq.}
\newcommand{\T}{Table}
\renewcommand{\S}{Sec.}

\newcommand{\ignore}[1]{}

\lstdefinestyle{base}{
  moredelim=**[is][\color{red}]{@}{@},
  escapeinside={<@}{@>}
}

\newcommand{\tool}{\textsc{SafeSCA}\xspace}

\newcommand{\parh}[1]{\noindent\textbf{#1}}

\usepackage{amssymb}
\usepackage{pifont}

\newcommand\DejaVuttfamily{%
  \fontfamily{DejaVuSansMono-TLF}\selectfont }

\lstdefinestyle{base}{
  moredelim=**[is][\color{red}]{@}{@},
  escapeinside={<@}{@>}
}

\lstdefinelanguage
   [x64]{Assembler}     
   [x86masm]{Assembler} 
   {morekeywords={CDQE,CQO,CMPSQ,CMPXCHG16B,JRCXZ,LODSQ,MOVSXD, %
                  POPFQ,PUSHFQ,SCASQ,STOSQ,IRETQ,RDTSCP,SWAPGS, %
                  rax,rdx,rcx,rbx,rsi,rdi,rsp,rbp, %
                  r8,r8d,r8w,r8b,r9,r9d,r9w,r9b}} 

\renewcommand{\baselinestretch}{0.980}

\usepackage{listings}
\usepackage{fancyvrb}
\usepackage{color}
\definecolor{lightgray}{rgb}{.9,.9,.9}
\definecolor{darkgray}{rgb}{.4,.4,.4}
\definecolor{purple}{rgb}{0.65, 0.12, 0.82}
\definecolor{commentgreen}{RGB}{63,127,95}

\colorlet{myPurple}{blue!40!red}
\definecolor{myOrange}{RGB}{255,192,0}

\newcommand{\enc}[1]{$\phi^{*}_{\theta}$}
\newcommand{\dec}[1]{$\psi^{*}_{\theta}$}

\lstdefinelanguage{Solidity}{
  keywords={len,delete,int,void,payable, public, event, contract, typeof, new, true, false, catch, function, return, null, catch, switch, var, if, in, while, do, else, case, break,struct,const,socklen_t,sa_familty_t,char,sockaddr},
  keywordstyle=\color{violet}\bfseries,
  ndkeywords={class, export, boolean, throw, implements, import, this},
  ndkeywordstyle=\color{darkgray}\bfseries,
  identifierstyle=\color{black},
  sensitive=false,
  comment=[l]{//},
  escapeinside={(*@}{@*)},          
  morecomment=[s]{/*}{*/},
  commentstyle=\color{commentgreen}\ttfamily,
  stringstyle=\color{red}\ttfamily,
  morestring=[b]',
  morestring=[b]"
}

\newcommand{\rnum}[1]{\uppercase\expandafter{\romannumeral #1\relax}}

\usepackage{xcolor}

\algnewcommand{\LeftComment}[1]{\Statex \(\triangleright\) #1}

\definecolor{pptbrown}{RGB}{132,60,12}
\definecolor{pptgreen}{RGB}{56,87,35}

\let\OLDthebibliography\thebibliography
\renewcommand\thebibliography[1]{
  \OLDthebibliography{#1}
  \setlength{\parskip}{0pt}
  \setlength{\itemsep}{0pt plus 0.1ex}
}

\definecolor{pptgreen}{RGB}{84,130,53}
\definecolor{pptred}{RGB}{176,35,24}
\definecolor{pptblue}{RGB}{194,214,236}

\definecolor{pptgreen1}{RGB}{78,173,91}
\definecolor{pptred1}{RGB}{192,0,0}

\definecolor{pptyellow1}{RGB}{203,195,167}
\definecolor{pptgreen2}{RGB}{184,192,176}

\newcommand{\bat}{\texttt{BAT}\xspace}

\newcommand{\bsfinder}{B2SFinder\xspace}

\newcommand{\centris}{\textsc{Centris}\xspace}

\newcommand{\binaryai}{BinaryAI\xspace}

\newcommand{\gemini}{\textsc{Gemini}\xspace}

\newcommand{\safe}{\textsc{SAFE}\xspace}

\newcommand{\crypten}{CrypTen\xspace}
\newcommand{\tplite}{TPLite\xspace}
\newcommand{\ossfp}{OSSFP\xspace}

\DeclareRobustCommand{\authororg}[1]{\smash{\textsuperscript{\footnotesize #1}}}
\begin{document}

\title{\fontsize{23.5}{10}\selectfont Preserving Privacy in Software Composition
Analysis: A Study of Technical Solutions and Enhancements}

\author{
    \IEEEauthorblockN{Huaijin Wang\authororg{1}, Zhibo Liu\authororg{1}\IEEEauthorrefmark{2}, Yanbo Dai\authororg{2},
    Shuai Wang\authororg{1}, Qiyi Tang\authororg{3}, Sen Nie\authororg{3}, Shi Wu\authororg{3}}
    \IEEEauthorblockA{\small \authororg{1}~Hong Kong University of Science and Technology,
    \authororg{2}~Hong Kong University of Science and Technology (Guangzhou),\\
    \authororg{3}~Keen Security Lab, Tencent\\
    \texttt{\{hwangdz, zliudc, shuaiw\}@cse.ust.hk,
    ydai851@connect.hkust-gz.edu.cn,
    \{dodgetang, snie, shiwu\}@tencent.com}}
}

\maketitle

\begin{abstract}
Software composition analysis (SCA) denotes the process of identifying
open-source software components in an input software application.
SCA has been extensively developed and adopted by academia and industry.
However, we notice that the modern SCA techniques in industry scenarios still
need to be improved due to \textit{privacy} concerns. Overall, SCA requires
the users to upload their applications' source code to a remote SCA server,
which then inspects the applications and reports the component usage
to users. This process is privacy-sensitive since the applications may
contain sensitive information, such as proprietary source code, algorithms, trade
secrets, and user data.

Privacy concerns have prevented the SCA technology from being used in
real-world scenarios. Therefore, academia and the industry demand
privacy-preserving SCA solutions. For the first time, we analyze the privacy
requirements of SCA and provide a landscape depicting possible technical
solutions with varying privacy gains and overheads. In particular, given
that de facto SCA frameworks are primarily driven by code similarity-based
techniques, we explore combining several privacy-preserving protocols to
encapsulate the similarity-based SCA framework. Among all viable solutions,
we find that multi-party computation (MPC) offers the strongest privacy guarantee
and plausible accuracy; it, however, incurs high overhead ({$184\times$}).
We optimize the MPC-based SCA framework by reducing the amount of crypto
protocol transactions using program analysis techniques. The evaluation
results show that our proposed optimizations can reduce the MPC-based SCA
overhead to only 8.5\% without sacrificing SCA's privacy guarantee or accuracy.
\end{abstract}

\let\thefootnote\relax\footnotetext{{\IEEEauthorrefmark{2} Corresponding author.}}

\vspace{-2pt}
\section{Introduction}
\label{sec:intro}

Given a piece of software, software composition analysis (SCA) identifies
components borrowed from third-party and open-source software (OSS) projects.
Hence, according to the component-usage information, developers are aware of the
usage of vulnerable or outdated OSS projects and, therefore, are able to take
further actions to reduce security risk, ensure license compliance, and promote
healthy open-source project usage. In recent years, the
industry~\cite{synk,whitesource,blackduck,codesentry} and
academia~\cite{sourcerercc,centris,wukong,libscout,libradar,libpecker,libd,bsa2bsca2024,
li2018large,libid,atvhunter,chen2020machine,chen2020automated} have extensively
studied SCA techniques in different settings and achieved promising results
regarding accuracy and efficiency.

Despite the prosperous development and adoption of SCA techniques, we notice
that the accompanying \textit{privacy concerns} are hindering the usage of SCA in
real-world scenarios. Indeed, some authors of this paper have been
working in a leading industrial company of SCA solutions, and they are frequently
encountering and handling requests for performing SCA without disclosing
customer's data in recent years. The typical pipelines of industrial SCA
solutions require a customer to upload the software to a remote server
possessed by the SCA vendor, where the software is analyzed to
produce a report that reveals the potential security and legal risks of the
software.
The report usually contains a list of reused OSS
projects~\cite{binaryaiurl,binaryai2024,sbom},
some of which may include known vulnerabilities.
This process is
privacy-sensitive since the uploaded software may contain sensitive
information, such as trade secrets or personal data. Moreover, the source
code is proprietary and deemed part of the customer's intellectual property (IP).
In reality, few companies are willing to provide commercial products' source
code to another party since it is hazardous to leak valuable IPs~\cite{wang2024pp}.

To avoid leaking customers' privacy, contemporary SCA vendors may deploy the SCA
service in the customer's in-house domain. While this alleviates the privacy
concerns from the user side, it requires the SCA vendor to share its IPs
(i.e., the SCA frameworks, algorithms, and accompanying data) with the customers.
For instance, When the SCA service is deployed on the customer's side, the SCA
vendor has to release the OSS database to customers periodically.
However, existing SCA vendors are often unwilling to share the 
valuable OSS database, which requires many resources to build.
Existing works~\cite{tplite2023issta,centris}
show that a comprehensive OSS
database with reliable OSS dependencies has an essential impact on the accuracy
of SCA, and it is not economical for small and medium-sized companies to build
their own databases. For instance, it is disclosed that
\binaryai~\cite{binaryaiurl}, an industrial SCA vendor, has spent over {three
years}, a dozen (senior) engineers, and PB-level disk space to construct its
OSS database. In addition, \binaryai relies on a large model that has been
trained with ten Nvidia Tesla T4 GPUs for years. Thus, exposing the model's
parameters is unacceptable.

Since SCA serves as one of the cornerstones of software security and a paramount
part of software engineering, the industry and academia are demanding a
privacy-preserving SCA solution to promote the practical adoption of SCA in
industry scenarios. Accordingly, this paper targets this practical need to study
the possible technical solutions for privacy-preserving SCA, and
we aim to answer the following two research questions (RQs):

\begin{itemize}
    \vspace{-5pt}
    \setlength\itemsep{0.1em}
    \item \textbf{RQ1:} How can we effectively protect both customers' and
    vendors' privacy in SCA?
    \item \textbf{RQ2:} How can a privacy-preserving SCA pipeline be built with
    satisfactory performance for practical use?
    \vspace{-5pt}
\end{itemize}
To answer \textbf{RQ1},
we systematically study existing technical
solutions and their (dis)advantages, exploring and proposing two novel
privacy-preserving SCA frameworks. As shown in \T~\ref{tab:whole-pic}, each
studied solution exhibits distinct privacy guarantees, required
resources, overhead, and potential pitfalls. Furthermore, we concretize the technical
solutions and empirically benchmark their overheads and accuracies. Specifically,
the two novel solutions we proposed involve advanced
privacy-preserving protocols to offer privacy-preserving SCA.
One solution relies on similarity-based bucketization (SBB)~\cite{sbb2022},
a scale private similarity testing technique;
the other depends on multi-party computation (MPC)~\cite{yao1986generate,
bendlin2011semi,damgaard2012multiparty}, which enables multiple parties
to jointly perform computation without revealing their private data to each other.
\S~\ref{subsec:sbb-sca-leakage} demonstrates that the SBB-based
solution, although lightweight, still bears the risk of potential privacy leakage.
In contrast, MPC offers a much stronger privacy guarantee by
shielding both the SCA customers' and vendors' valuable assets,
such as code, model, and data.
Overall, our study illustrates the promising potential
of MPC-based SCA.

Nevertheless, the MPC-based SCA pipeline we proposed incurs excessively 
high overhead ($184\times$), which prohibits its practical usage.
To answer \textbf{RQ2},
we propose three optimization strategies with program analysis techniques,
which significantly reduce the amount of crypto protocol transactions.
Our optimization strategies, including (1) \textit{symbol filter},
(2) \textit{informative source function selector}, and (3)
\textit{assembly function selector},
can vastly reduce the time cost of MPC-based SCA from $184\times$ to $23\times$,
and the most time-consuming process, i.e., encrypted computation with MPC,
is reduced to 4.8\%,
leading to a practical privacy-preserving SCA solution.
In sum, this work presents the following contributions:
\vspace{-5pt}
\begin{itemize}
    \setlength\itemsep{0.2em}
    \item Observing the privacy concerns of SCA over both the user and the
    SCA service provider, we, for the first time, study the privacy-preserving SCA,
    whose absence is a crucial hindrance to the practical adoption of SCA in
    industry scenarios.
    \item We analyze the privacy leakage risks in three typical SCA
    solutions and formulate two novel privacy-preserving SCA frameworks with SBB
    and MPC, respectively. Each studied solution exhibits distinct
    privacy guarantees and required resources. We then present an
    empirical evaluation of their overheads and accuracy.
    We find that MPC offers high accuracy with the strongest privacy guarantee.
    \item We design a set of optimizations that can largely
    reduce the cost of expensive crypto operations for the MPC-based SCA
    solution and without sacrificing accuracy,
    achieving the first practical privacy-preserving SCA.
\end{itemize}
\vspace{-5pt}
\parh{Artifact Availability.}~We have released our artifact at~\cite{snapshot}.
We will maintain it for future research comparison and usage.

\vspace{-3pt}
\section{Background}
\label{sec:bkg}

\begin{figure}[]
  \vspace{-5pt}
  \includegraphics[width=1.0\linewidth]{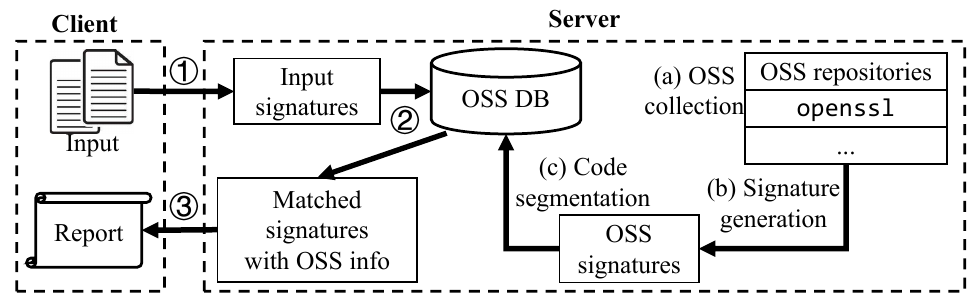}
  \vspace{-15pt}
  \caption{A common SCA pipeline.}
  \label{fig:sca}
  \vspace{-20pt}
\end{figure}

\vspace{-3pt}
\subsection{A Common SCA Pipeline}
\label{subsec:common-sca}
This section introduces the common SCA pipeline, which is widely adopted by
state-of-the-art (SoTA) C/C++ SCA works~\cite{centris,
wu2023ossfp,tplite2023issta,yang2022modx,libdb2022,binaryaiurl}. As shown in
\F~\ref{fig:sca}, the pipeline comprises a server (usually owned by
the SCA vendor) and a client for customers to access the SCA service.
To perform SCA, the server first builds an OSS database (OSS DB)
offline (i.e., (a), (b), and (c)) and then provides an online SCA service (i.e.,
edges \ding{172}, \ding{173}, and \ding{174}) to the client.
Accurate SCA analysis relies on extracting high-quality distinguishable
code signatures that can help identify semantic differences between code snippets.
Recent SCA works~\cite{centris,atvhunter,tplite2023issta,bsa2bsca2024,
libdb2022,sourcerercc} generate signatures with advanced similarity analysis
techniques such as locality-sensitive hashing (LSH)~\cite{wu2023ossfp,centris}
and deep learning-based embeddings~\cite{libdb2022,binaryai2024,atvhunter} to
extract features from source or binary code.
Below, we introduce the major steps in SCA. 
Moreover, \S~\ref{subsec:centris-tlsh} discusses a concrete example, \centris.

\smallskip
\parh{Offline Analysis.}~The offline process for building OSS DB
consists of three steps: \textit{OSS collection}, \textit{signature generation},
and \textit{code segmentation}.

\noindent \textit{OSS Collection.}~An extensive OSS DB is inevitable for real-world
SCA tasks. Existing SCA techniques (e.g., \centris~\cite{centris} and
\tplite~\cite{tplite2023issta}) build their OSS DB by downloading reputed
repositories from open-source platforms like GitHub and GitLab.
In this work, we build our OSS database from the same sources. More details are discussed in
\S~\ref{subsec:oss-database}.

\noindent \textit{Signature Generation.}~The SCA server dissects an OSS
repository into smaller code pieces with finer granularities (e.g.,
functions and source files). Then, the server generates signatures for each
code piece based on specific underlying SCA techniques.
A common practice is to generate function-level signatures with
hash algorithms~\cite{wu2023ossfp,centris} or embedding
techniques~\cite{libdb2022,binaryai2024,atvhunter}.
Therefore, each OSS repository in the OSS DB has a set of function-level
signatures (e.g., hash values or embedding vectors).

\noindent \textit{Code Segmentation.}~A common observation is that an OSS project often
reuses other OSS projects. Therefore, the signatures of an OSS project may
overlap with some other projects' signatures due to the borrowed code. For
instance, \texttt{SDL}~\cite{sdlurl},
a popular media library,
borrows code from \texttt{yuv2rgb}~\cite{yuv2rgburl},
an image conversion library. Thus, some function
signatures of \texttt{yuv2rgb} are also included in \texttt{SDL}'s
signatures.
The overlapped signatures likely result in false positives in SCA; e.g., a
project reusing \texttt{yuv2rgb} is likely to be identified as reusing
\texttt{SDL}. Existing works (e.g., \centris~\cite{centris} and
\tplite~\cite{tplite2023issta}) segment the sets of OSS projects' signatures
into non-overlapped sets to address this issue.

\smallskip
\parh{Online Analysis.}~To conduct SCA, the client first
uploads the input source or binary code to the server and generates signatures
derived from the input (\ding{192}). Then, the server searches for similar or
identical signatures in the OSS DB and records the matched signatures with
their OSS information (\ding{193}). Finally, the server returns the matched
information and the SCA report to the client
(\ding{194}). Details of the three steps are below.

\noindent \ding{192} The client sends the target software's codebase (e.g.,
source or binary code) to the SCA server, which computes the input signatures
in the same way as the offline signature generation. Precisely, the server
dissects the input codebase into smaller code pieces (e.g., functions) and then
generates signatures (with hash or embedding models) for each code piece.

\noindent \ding{193} Given the input signatures, the server searches for relevant
signatures in the OSS DB. Since signatures generated from similar code pieces
have short distances,
if an input signature is similar or identical to one signature in the OSS DB,
the code piece is likely borrowed from the corresponding OSS project.

\noindent \ding{194} Having identified which code snippets are being borrowed
from OSS projects, the server reports the SCA results to the client, which
depict the information on reused OSS projects. Specifically, the server can warn
the client if the reused OSS projects have known vulnerabilities or license
issues.

\vspace{-3pt}
\subsection{\centris (TLSH)}
\label{subsec:centris-tlsh}
To better understand the pipeline, we take CENTRIS~\cite{centris}, a SoTA
source-based SCA framework depending on the Trend Micro Locality Sensitive Hash
(TLSH)~\cite{oliver2013tlsh} for signature generation and OSS identification, as
an example. For simplicity, we refer to it as ``\centris (TLSH)'' to distinguish
it from other \centris variants.

\smallskip
\parh{TLSH.}~TLSH is a locality-sensitive hash algorithm designed for
similarity testing.
TLSH generates a hash value for one function and ensures similar functions
result in close hash values. In practice, if the distance between two hash
values is smaller than a threshold ($\delta = 30$), the two source functions are
considered from the same source with high confidence. Specifically, with the
distance measure function ($dis$), the functions $f_i$ and $f_j$ are similar if
their hash values $h_i$ and $h_j$ satisfy $dis(h_i, h_j) < \delta$.

\smallskip
\parh{Offline Analysis.}~Given the set of collected OSS repositories
$O = \{o_1, o_2, \dots, o_n\}$, \centris first dissects $o_k \in O$
into a set of source functions $F_k$, and then generates TLSH hash values
for each function in $F_k$, which forms the function hash value set $H_k$.
After processing all OSS repositories, for each OSS $o_k$,
\centris segments the hash value
set $H_k$ into a non-overlapped set $DB_k$, which
is stored in the OSS DB for online analysis.

\smallskip
\parh{Online Analysis.}~
Assume there are $n$ OSS projects in the OSS DB, and the function hash value set
of the OSS project $k$ is $DB_k$. First, given the target software's
codebase, \centris dissects it into source functions and generates TLSH hash
values for each function, resulting in the function hash value set $I$. 
Then, \centris searches for similar hash values and their original OSS projects
in the OSS DB. For each OSS project $k$, \centris checks the set of similar hash
values between $I$ and $DB_k$ with \E~\ref{eq:centris1} and gets the set of
similar function pairs $S_k$.
Finally, if $ \frac{|S_k|}{|DB_k|} $ is greater than a threshold
$\epsilon$ (set as 0.10 in \centris),
which denotes that a certain portion of functions are
deemed as reused, \centris reports the OSS project $k$ as reused. By
traversing all OSS projects, CENTRIS can report all reused OSS projects in the
input codebase.
\begin{equation}
  \hspace{-3pt} S_k = \{(h_i, h_j) | \forall (h_i, h_j) \in I \times DB_k\ \text{and}\ dis(h_i, h_j) < \delta\}
  \hspace{-3pt} \label{eq:centris1}
\end{equation}

\vspace{-3pt}
\subsection{Privacy-Preserving Protocols}
\label{subsec:ppp}

Privacy-preserving protocols allow users to protect their privacy
while maintaining the service without being
affected. In this paper, privacy-preserving protocols aim to protect SCA
consumers' and vendors' IPs from being exposed to other parties
while allowing a valid and accurate SCA analysis. 
This section introduces the relevant mainstream privacy-preserving protocols
that are relevant to our work.

\begin{figure}
  \vspace{-5pt}
  \centering
  \includegraphics[width=0.9\linewidth]{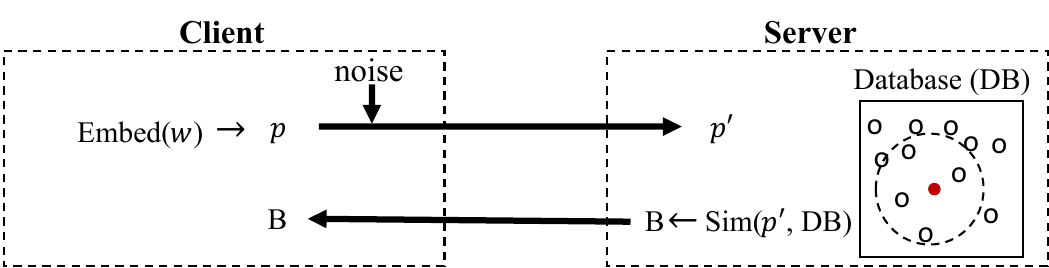}
  \vspace{-3pt}
  \caption{Similarity-based bucketization protocol.}
  \label{fig:sbb}
  \vspace{-15pt}
\end{figure}

\smallskip
\parh{SBB.}~Similarity-based bucketization (SBB)~\cite{sbb2022}
provides private similarity testing. Compared with expensive
cryptographic protocols, SBB scales well to large databases and does not
sacrifice the correctness of matching. \F~\ref{fig:sbb} presents the framework
of SBB. Given a client's element ($w$) to be queried, the client first
embeds the element and get its representation $p$. After mutating $p$ with
noises, the mutated result $p'$ is sent to the server. Then, the server
searches for similar elements with $p'$ (i.e., the \textcolor{red}{red} point) in
its database and get a bucket (i.e., B) of similar elements, which is then
returned by the server. After receiving the bucket, the client searches for the
most similar element with the non-mutated $p$.
Since the server can only get the mutated $p'$, the client's privacy is
preserved.
\S~\ref{subsec:sbb-sca} will discuss the application of SBB in the SCA scenario.
Nevertheless, the mutated $p'$ may still leak some information to the 
SCA vendor. \S~\ref{subsec:sbb-sca-leakage} will discuss the SBB leakage in
the SCA scenario.

\smallskip
\parh{MPC.} Multi-party computation (MPC) is a
privacy-preserving paradigm that allows participating parties to jointly perform
computations over each party's local data. Following exact MPC protocols,
nothing except for the correct computation result is revealed to each party.
Practical MPC protocols can be either garbled-circuits-based~\cite{yao1986generate}
or secret-sharing-based methods~\cite{bendlin2011semi,damgaard2012multiparty,
crypten}. We use \crypten~\cite{crypten}, a secret-sharing method for machine
learning, to implement our framework.

SCA involves two parties, including the client and the server. The secret is
split into two shares, one for each party, and the secret cannot be constructed
without the agreement between the two parties. To facilitate secure
computations, \crypten implements arithmetic secret sharing and binary secret
sharing. In such cases, private addition between two arithmetically secret
shared values, $[z]= [x] + [y]$, can be implemented by having each party locally
sum their shares of $[x]$ and $[y]$. Specifically, each party $p\in\mathcal{P}$
needs to locally compute $[z]_p= [x]_p + [y]_p$.
Private multiplication between two arithmetically secret shared values is
implemented with the assistance of random Beaver triples \cite{beaver1992efficient}, which
consists of $([a],[b],[c])$ with $c=ab$, provided by the trusted third party.
Each party computes $[\epsilon]=[x]-[a]$ and $[\delta]=[y]-[b]$ and then
reconstructs $\epsilon$ and $\delta$ through exchanging the shares. The result
can then be computed $[x][y]=[c]+\epsilon[b]+\delta[a]+\epsilon\delta$.

While the application of MPC protocols provides cryptographically guaranteed
privacy, they often introduce significant overhead, which is influenced by the
number of parties and the complexity of the model. There are always two parties
in our scenario, and the model remains unchanged after encryption.

\section{Privacy Concerns in SCA and Motivation}
\label{sec:motivation}

In this section, we discuss the assets to be protected in SCA
(\S~\ref{subsec:privacy-concerns}),
the leakage of existing SCA techniques (\S~\ref{subsec:leakage-sca}),
and the motivation for proposing privacy-preserving SCA.
In the following
discussion, we assume the customer and the vendor hold the client and the server
in \F~\ref{fig:sca}, respectively, unless otherwise stated (i.e., private
deployment).

\vspace{-3pt}
\subsection{Assets of Concern in SCA}
\label{subsec:privacy-concerns}

\parh{Valuable Assets.}~As shown in \F~\ref{fig:sca},
the online SCA service involves three resources, including
the client's input (in source or binary code format), the SCA report,
and the server's OSS database (with any embedding models used to
establish the database).
We consider those resources (listed in \T~\ref{tab:whole-pic})
as valuable assets to be protected for privacy security.
As aforementioned in \S~\ref{sec:intro}, the privacy of customers' code and
vendors' OSS database and model are essential.
As the importance of protecting customers' source code
and the vendor's model is self-evident,
and securing the vendor's database has been discussed in
\S~\ref{sec:intro} (e.g., \binaryai~\cite{binaryaiurl,binaryai2024}
keeps the database private),
we discuss the necessity of protecting the client's binary code and
SCA reports below.

\smallskip
\parh{Privacy Concerns for Binary Code.}~Although the compiled binary
code is less comprehensible than the source code, it still contains
sensitive information that can be reverse-engineered and disclosed.
Therefore, customers are unwilling to share binaries 
due to potential security and IP risks~\cite{manadhata2004measuring,manadhata2010attack}.
For instance,
leaking an online game's server executable results in deploying private servers
easily and causing catastrophic financial loss~\cite{maplestorysuite}. Even if a
binary is publicly available, developers may only release the secured binaries
to protect their IP by hindering reverse engineering~\cite{tigress,rlobf2020,
junod2015obfuscator,ding2019asm2vec,luo2014semantics,luo2017semantics}. For
instance, many Android apps are packed and obfuscated before release. Since
reverse engineering on those binaries is challenging, binary-based SCA works can
hardly extract accurate features~\cite{libdb2022}, resulting in unreliable SCA
reports.
Thus, it is necessary to consider binary code as a valuable asset to be protected.

\smallskip
\parh{Privacy Concerns for SCA Reports.}~Leaking the SCA report allows a
potential attacker to know what OSS is used by the customer; thus, if the
customer uses an OSS with a known vulnerability (or weakness), the attacker can
directly exploit it instead of searching for vulnerabilities blindly, which is
often a time-consuming step in real-world attacks~\cite{crosby2003denial,
praseed2018ddos}. Hence, leaking the SCA report can ease the attacker's effort
in searching and exploiting software vulnerabilities. For example, once an
attacker knows a web server uses the vulnerable Log4j~\cite{log4jurl}, the
attacker can reuse the disclosed attack resources (e.g., crafted payloads) for
other systems depending on the vulnerable Log4j to attack the target web server.

\begin{table}[]
  \caption{Leaked resources of various settings. \leaked\ denotes resources are
  leaked. ``NA'' means the SCA solution does not rely on a trained model.}
  \label{tab:whole-pic}
  \vspace{-5pt}
  \hspace{-10pt}
\setlength{\tabcolsep}{3pt}
  \resizebox{1.00\linewidth}{!}{
  \begin{tabular}{c|ccc|cc|c}
      \hline
      \multirow{3}{*}{SCA Solution}   & \multicolumn{5}{c|}{Valuable Assets} & \multirow{3}{*}{Tool} \\ \cline{2-6}
                          & \multicolumn{3}{c|}{Customer-side} & \multicolumn{2}{c|}{Vendor-side} &  \\ \cline{2-6}
                          & Src & Bin & Report & Model & DB & \\ \hline
      Source-based    & \leaked & \secure & \leaked &         NA & \secure & \centris (TLSH) \\ 
      Private deployment  & \secure & \secure & \secure &         NA & \leaked & \centris (TLSH) \\ 
      Binary-based    & \secure & \leaked & \leaked & \secure    & \secure & \binaryai \\ \hline
      SBB-based       & \secure & \secure & \leaked &         NA & \leaked & \centris (TLSH + SBB)     \\ \hline
      \multirow{2}{*}{MPC-based}  & \multirow{2}{*}{\secure} & \multirow{2}{*}{\secure} & \multirow{2}{*}{\secure} & \multirow{2}{*}{\secure}    & \multirow{2}{*}{\secure} & \centris      \\ 
                 &         &         &         &            &         & (DPCNN + \crypten)      \\
    \hline
  \end{tabular}
  }
  \vspace{-15pt}
\end{table}

\vspace{-2pt}
\subsection{Existing SCA Solutions}
\label{subsec:leakage-sca}

Based on the input types, we categorize existing SCA techniques into
\textit{source-based SCA} and \textit{binary-based SCA}. Besides, we also
observe that some companies deploy the server of source-based SCA on the
customer side (we refer to it as \textit{private deployment}) to protect
customers' proprietary source code. These representative SCA solutions and their
privacy leakage are summarized in \T~\ref{tab:whole-pic}.

Assume a customer develops a software project $P$; the source code of $P$ is
$P_{src}$, and the compiled (and stripped) binary is $P_{bin}$. The vendor owns
an OSS database $DB$ and provides an SCA service to detect the reused OSS in
$P$. 
Besides, binary-based SCA may use an embedding model $M$ for
signature generation due to the difficulties in analyzing binaries~\cite{binaryai2024}.
Initially, the customer owns $P_{src}$ and $P_{bin}$,
and the vendor owns $DB$ and $M$. After the analysis, the customer
receives an SCA report $R$.
When $P_{src}$, $P_{bin}$, and $R$ are available to the vendor,
we mark them as \leaked\ in \T~\ref{tab:whole-pic}.
The vendor's OSS database $DB$ and model $M$ are marked as \leaked\
if they are leaked to the customer.
The three SCA solutions and their leakages are discussed separately below.

\smallskip
\parh{Source-Based SCA.}~To perform SCA,
the customer's client uploads $P_{src}$ to the vendor's server,
which then generates the SCA report $R$ and sends it back to the client.
While OSS DB is not disclosed to the customer's client,
the SCA server can access $P_{src}$ and $R$.
Therefore,
the customer's $P_{src}$ and $R$ are leaked, while the vendor's $DB$
is secure.

\smallskip
\parh{Private Deployment (of Source-Based SCA).}~To address the customer's
privacy concerns,
some vendors deploy the SCA server on the customer side.
In this way, the customer has full control over the SCA client and
server, while the vendor needs to provide the OSS database $DB$.
Hence, the customer's $P_{src}$, $P_{bin}$, and $R$ are secure,
but the vendor's $DB$ is leaked.

\smallskip
\parh{Binary-Based SCA.}~The customer's binary
$P_{bin}$ is uploaded to the vendor's server, while the source code $P_{src}$ is
not disclosed. After receiving $P_{bin}$, the vendor performs reverse
engineering and SCA analysis, generates the SCA report $R$, and sends it back to
the customer. As a result, the vendor has access to $P_{bin}$ and $R$, while its
OSS database $DB$ and model $M$ for signature generation are secure.

\parh{Source-Based SCA vs. Binary-Based SCA.}~Binary-based SCA is mainly
proposed for scenarios where source code is unavailable to the vendor, which
can be caused by the usage of closed-source
libraries or by the customer's privacy concerns.
Compared with source-based SCA, binary-based SCA only leaks obscure
binary code to the vendor. However, as discussed earlier, a binary still exposes
sensitive information when it is reverse-engineered. On the other hand, binary
analysis is difficult~\cite{libdb2022,b2sfinder,liu2022sok,liu2020far,liu2023decompiling,
ding2019asm2vec,wang2022jtrans,binaryai2024,wong2022deceiving}, leading to a relatively low SCA
accuracy. In \S~\ref{subsec:performance-sca}, we show that
\binaryai~\cite{binaryaiurl,binaryai2024}, a commercial
binary-based SCA tool, has relatively
low accuracy compared with source-based techniques. We thus deem binary-based
SCA is far from an ideal privacy-preserving solution.
\vspace{-2pt}
\begin{tcolorbox}[size=small]
  \parh{Motivation.}~Existing SCA solutions have different privacy concerns.
  Source- and binary-based SCA leaks customers' assets, while private deployment
  leaks vendors' assets.
  Thus, there is a dilemma between customers' and vendors' privacy in SCA,
  and addressing the dilemma with a privacy-preserving SCA solution is an urgent need.
\end{tcolorbox}
\vspace{-2pt}


\section{Privacy-Preserving SCA Frameworks}
\label{sec:pp-sca}
This section illustrates the privacy-preserving SCA frameworks, where
we expect that neither the client nor the server can learn the other's valuable
assets (see \S~\ref{subsec:privacy-concerns}). Specifically, the vendor cannot
learn the client's source code, binary code, and the SCA report, while the
client cannot learn the vendor's database and model (if any). 
Before introducing their designs in detail, we first clarify the threat model.

\smallskip
\parh{Threat Model.}~We consider a \textit{semi-honest} threat model, where the
client and the server are assumed to follow the protocol specification
exactly~\cite{lindell2017simulate,sbb2022}. However, they are curious about each
other's private information and record all intermediate
results~\cite{goldreich2004foundations}. We assume the customer and the SCA
vendor are entities with a reputation and are not likely to conduct maliciously
adversarial behaviors.

In order to design privacy-preserving SCA frameworks, we adapt the
privacy protocols to an existing SoTA SCA solution.
Specifically, we enhance \centris~\cite{centris},
which is introduced in \S~\ref{subsec:centris-tlsh}.
We select advanced and representative privacy-preserving protocols
to protect the privacy of the SCA pipeline in two aspects:
(1) protecting the signature matching process based on similarity
algorithm (i.e., \ding{193} in \F~\ref{fig:sca})
and (2) encrypting the package transactions between the client and the server (i.e.,
\ding{192} and \ding{194} in \F~\ref{fig:sca}).

We tentatively adapt SBB to protect the signature matching process, resulting in
an SCA framework referred to as ``\centris (TLSH + SBB)'', whose details are
presented in \S~\ref{subsec:sbb-sca}.
However, we notice that such a solution leaves SCA reports
and OSS DB unprotected (explained in \S~\ref{subsec:sbb-sca-leakage}),
we resort to the other
approach that employs MPC to protect the privacy of different
parties with encryption.
Specifically, we enhance \centris with code embedding
techniques~\cite{henkel2018code,word2vec2013,Word2Vec}
(referred to as ``\centris (DPCNN)'' in
\S~\ref{subsec:centris-dpcnn}) and then adapt the popular MPC framework,
\crypten~\cite{crypten}, to protect it (referred to as ``\centris
(DPCNN + \crypten)'' in \S~\ref{subsec:mpc-sca}).

\begin{figure}
    \centering
    \includegraphics[width=1.00\linewidth]{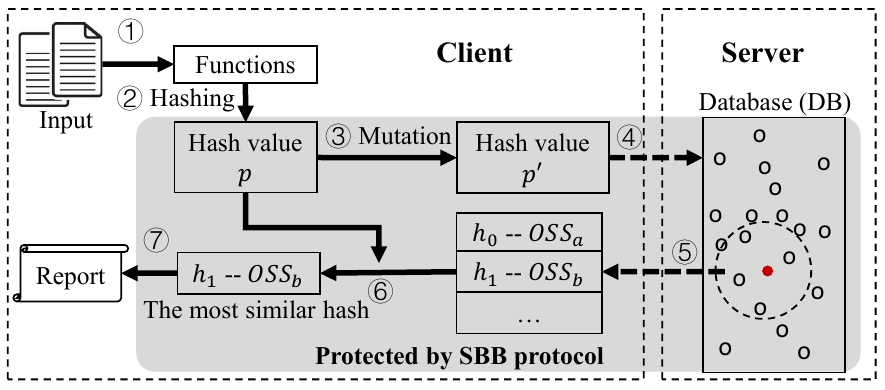}
    \vspace{-20pt}
    \caption{Workflow of SBB-based SCA.}
    \label{fig:sbb}
    \vspace{-15pt}
\end{figure}

\subsection{SBB-Based SCA (\centris (TLSH + SBB))}
\label{subsec:sbb-sca}

\F~\ref{fig:sbb} presents the workflow of SBB-based SCA.
The client first dissects the input source repository into
functions (\ding{192}) and then generates the signatures
with TLSH (\ding{193}). By employing the SBB protocol to
protect the private data, a function's hash value $p$ will not
be sent to the server directly. Instead, the client
mutates the hash value $p$ into a new hash value $p'$ (\ding{194}),
and sends $p'$ to the server (\ding{195}).
The server then searches for similar signatures of $p'$ (i.e., the {\color{red}{red}}
point) in the OSS DB and replies to the client with
a bucket of signatures close to $p'$ (\ding{196}).
After receiving the bucket, the client can identify the most similar
signature in the bucket with $p$ (\ding{197}).
After repeating the process from \ding{193} to \ding{197} for all
functions, the client can report the SCA results to the customer (\ding{198}).

\smallskip
\parh{Hyperparameters of SBB.}~The SBB protocol has two hyperparameters: the
distance threshold $\gamma$ for mutating the hash value $p$ to $p'$, and the
bucket size $\theta$ for the server to reply to the client. Intuitively, if
$\gamma$ is too small, the mutated hash value $p'$ is close to $p$, which
discloses the information of the client's source code; however, if $\gamma$ is
too large, the server has to use a large $\theta$, which can result
in high overhead in identifying the most similar signature in the bucket
(\ding{197}); otherwise, the exact similar signatures of $p$ may be missed in
the replied bucket.

\smallskip
\parh{Privacy Concerns of SBB-Based SCA.}~As shown in \F~\ref{fig:sbb},
the SBB protocol protects the client's private data by mutating
the signatures before sending them to the server.
The client locally generates SCA results according
to the embeddings returned by the server.
Thus, the server has no access to the original
signatures of source code and SCA results.
However, the mutated signatures
are similar to the original ones and still
disclose some information about the client's source code.
In \S~\ref{subsec:sbb-sca-leakage}, we demonstrate that the server
can infer the client's SCA report, given only the mutated
signatures.
Moreover, because the server replies to the client with
a bucket of similar signatures
(i.e., thousands of signatures) for each query, 
the client can efficiently steal the server's OSS DB
by repeatedly querying the OSS DB.
As a result, we mark the customer's SCA report and
the server's OSS DB as leaked (\leaked) in \T~\ref{tab:whole-pic}.

\vspace{-2pt}
\subsection{\centris (DPCNN)}
\label{subsec:centris-dpcnn}
Since there is no MPC solution for TLSH, we design a variant of
\centris (TLSH) that replaces TLSH with an embedding model, which
is a deep parallel convolutional neural network (DPCNN)~\cite{johnson2017deep}
that maps a source function to an embedding.
We use the Euclidean distance between two embeddings to evaluate
the similarity between two functions.

As mentioned in \S~\ref{subsec:centris-tlsh},
\centris (TLSH) identifies an OSS project as reused when 10\% of its
functions are recognized in the input. An observation is that the
threshold is too high for many OSS projects~\cite{tplite2023issta}.
Hence, like many existing static analysis methods~\cite{codeql,joern,
bdbaurl,binaryaiurl,
DBLP:journals/tosem/WangWYSZXZ23}, \centris (TLSH) struggles to
avoid false negatives, and lowering the threshold
cause a significant increase in false positives~\cite{centris}.
To further explore unleashing the full potential and achieving higher accuracy,
we consider assigning weights to the source functions, following the idea of
BAT~\cite{bat2011} for weighted strings.

\smallskip
\parh{Weight Enhancements.}~Given a function $f$ with similar
functions in the OSS database,
let the number of OSS repositories with similar functions be $N$,
and the lines of $f$ be $LoC$. We define the weight
of $f$ as $w = \frac{LoC}{5^{N-1}}$. The intuition behind this
formula is that a function with many similar functions is less
representative (e.g., helper functions) and should consequently have a lower weight.
On the other hand, a longer function usually contains rich semantics and should have
a higher weight. We then score each OSS project in the OSS DB by
the sum of the weights of its similar functions.
If an OSS project's score exceeds the threshold
$\beta$, it is identified as reused. \T~\ref{tab:accuracy1} shows the result
when $\beta=100$, which is the threshold used by \bat. To efficiently search
similar embeddings, we use Faiss~\cite{johnson2019billion}, an efficient
similarity search library provided by Meta, to build an index for all
embeddings.

\begin{figure}
    \centering
    \includegraphics[width=0.95\linewidth]{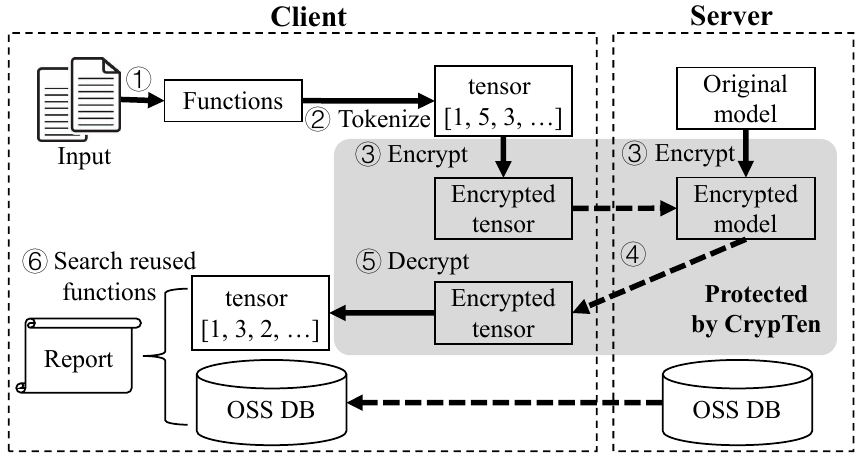}
    \vspace{-8pt}
    \caption{Workflow of MPC-based SCA.}
    \label{fig:mpc}
    \vspace{-15pt}
\end{figure}

\subsection{MPC-Based SCA (\centris (DPCNN + \crypten))}
\label{subsec:mpc-sca}
To adapt MPC to SCA, we build a SCA framework whose workflow is
presented in \F~\ref{fig:mpc}.
Compared with \centris (DPCNN), the data transaction between the client and
the server is protected by \crypten.
Before the online analysis, the client first downloads the OSS DB
from the server.
During online analysis, the client initially dissects the input
source repository into source functions (\ding{192}) and then tokenizes
and converts each function into a plaintext tensor (\ding{193}).
The client then encrypts the tensor with \crypten (\ding{194}),
and the server encrypts the original model with \crypten (\ding{194}).
Then, the client sends the encrypted tensor to the server for
encrypted computation and receives the encrypted results (\ding{195}).
After decrypting the results, the client gets the plaintext embedding
vectors of the functions (\ding{196}), with which
the client can search for similar embeddings in the OSS DB
and report the SCA results to the customer (\ding{197}).

\smallskip
\parh{Shared OSS Vector Database.}~To
avoid leaking the customer's privacy (e.g., SCA reports) to the SCA vendor,
the client needs to download the OSS DB from the vendor's server and query the DB locally.
Otherwise, if the vendor takes responsibility
for conducting DB queries, SCA reports will be revealed directly to
the vendor. As summarized in \T~\ref{tab:whole-pic},
source- and binary-based SCA have the same issue when the DB is
deployed on the vendor's server side.
Nevertheless, note that the OSS DB contains only embedded vectors (i.e.,
signatures) of OSS projects. Consequently, without the embedding model, which is
exclusive to the client, the client cannot exploit the DB or hurt the vendor's
privacy, which will be introduced in the next few paragraphs.
On the other hand, downloading the OSS DB requires additional storage
space and network bandwidth. We clarify that the cost is acceptable
since the client only downloads the OSS vector
database (around 10 GBs) instead of OSS repositories (over 800 GBs).

\smallskip
\parh{Privacy Concerns of MPC-Based SCA.}~From the client's perspective, the
\crypten protocol protects the client's private data by encrypting the data
before sending it to the server. Thus, the server cannot access the client's
source code and SCA results. Moreover, no information about the client's data is
leaked as the MPC protocol guarantees clients' privacy.

A notable concern from the server's perspective is that the server's OSS DB is
exposed to the client. However, we emphasize that the meaningful information of
the OSS database is \textit{still protected}. To clarify, the client uses the
OSS database to search for similar functions from OSS projects. Each record in
the OSS database stores the embedding vector of a function, and the information
about the project to which this function belongs. To use the OSS database, the
client needs access to the embedding model to embed functions for similarity
search. However, the model itself is \textit{not} disclosed to the client.
Consequently, even if the client has downloaded the OSS database, the client
still needs to request the server's service to use the database. In this way,
the only meaningful information accessible to clients is the set of included OSS
projects; therefore, the privacy of the OSS database is protected.

Another concern is that the client may approximate the server's plaintext model
with sufficient query results; we view this as \textit{orthogonal} and mostly
beyond the consideration of MPC-based model protection. To understand the
mitigation, we observe that the community has developed various techniques to
alleviate such model stealing attacks~\cite{li2022defending,kesarwani2018model,
juuti2019prada,flooding2021monitoring}, such as limiting the query frequency,
updating the model periodically, and detecting malicious queries. Second, in our
context, issuing sufficient queries is time-consuming due to the MPC protocol's
overhead. As a proof-of-concept, we studied stealing 1\% of the OSS database to
approximate the source model. The result is not promising (about 0.589 of the
plaintext model F1 score), and it takes about three days to finish all queries.
\vspace{-2pt}
\section{Benchmarking Privacy-Preserving SCA}
\label{sec:exp-ppsca}

In line with the landscape of privacy-preserving SCA solutions presented in
\S~\ref{sec:pp-sca}, this section performs empirical evaluations and discusses
the results. We first introduce the dataset and the OSS
database used in our evaluation and then present the evaluation results.
We additionally discuss the privacy leakage of the SBB-based SCA
at the end.

\vspace{-2pt}
\subsection{OSS Database Construction}
\label{subsec:oss-database}
Building a large-scale OSS database is inevitable for real-world SCA tasks.
Existing SCA techniques (e.g., \centris~\cite{centris} and
\tplite~\cite{tplite2023issta}) build their OSS database by downloading reputed
OSSs from open-source platforms like GitHub and GitLab. Since prior
works~\cite{centris,tplite2023issta} do not release their database or the
dependency ground truth due to privacy protection policies, we establish
a new large-scale database in this work. In particular, we build the
database by identifying all C/C++ software with GitHub URL references collected
from the CVE database. Initially, we collected 1,166 C/C++ software that was
referred by a CVE number. After that, we also cloned their Git submodules
recursively and removed repeated repositories. Eventually, we collected 1,832
C/C++ OSS repositories from GitHub. Those OSS projects have significantly
different sizes (ranging from 10 KB to 2.3 GB) and are developed for various
purposes (e.g., \texttt{NumPy} for scientific computing and \texttt{RocksDB} for
efficient database). Our database contains many more OSS
repositories than \tplite (1,036 repositories), demonstrating the
comprehensiveness of our setup.

\vspace{-2pt}
\subsection{Evaluation Dataset}
\label{subsec:dataset}

We pick 15 representative OSS projects, each with over 5 \texttt{Git}
submodules, from our collected OSS database as the evaluation dataset. We use
the \texttt{Git} submodules to ease the effort of marking the ground truth of
reused OSS projects. For each project, we compile it with \texttt{GCC}
with default compiler flags. We successfully built 14 projects
and formed a test dataset of 14 binaries.
The projects of 14 binaries contain about
37 million LoC with 211 identified OSS dependencies
in our OSS database.
Their diverse purposes include but are not limited to multimedia processing,
databases, drivers, and coin miners.

It is worth noting that the ground truth marked with \texttt{Git} submodules is
not perfectly accurate; we still need to manually check each project's source
and binary code due to the complex C/C++ software
ecosystem~\cite{tang2022towards}. Many OSS projects are reused in the form of
copy, and {there may be no announcement/documents detailing this in the original
OSS projects}. To build a reliable ground truth, we first tried to use \centris,
the SoTA source code-based SCA framework, to detect OSS reuses. However, the
results are unsatisfactory due to imperfect code segmentation and the
incapability of TLSH~\cite{oliver2013tlsh}
used by \centris. We then manually refine the results by searching for URLs,
keywords, file names, and representative function names in the source code.
Regarding directly reused binary code, we manually check libraries that are
statically linked to the binary. Conducting this manual check took two authors
about 40 man-hours. Each author has an in-depth knowledge of reverse
engineering, SCA, and rich experience in binary code analysis. This ensures the
credibility and accuracy of our dataset to a great extent.
Note that to avoid possible biases in the experiments and be close to the
real-world scenario, we remove these 14 OSS projects from our
OSS database.

\subsection{Experimental Setup}
\label{subsec:exp-setup}
In this section,
we elaborate on the setup of \centris and its variants,
and the selection of binary-based SCA tool (i.e.,
\binaryai~\cite{binaryai2024,binaryaiurl}).

\parh{\centris (TLSH).}~The official implementation of
\centris~\cite{centrisurl} is not sufficiently efficient to be applied to our
extensive OSS database since it searches close hash values in the OSS database
one by one. To accelerate the search process, we equip \centris\ with the HNSW index
algorithm~\cite{malkov2018efficient} for TLSH.
Other settings of \centris (TLSH) are
kept the same as the official implementation.
We present its performance in \T~\ref{tab:accuracy1}
as a ``lite'' SCA solution that can be deployed on a customer's side
due to its limited requirement for computation resources and public
availability.

\parh{\centris (TLSH + SBB).}~As described in \S~\ref{subsec:sbb-sca}, we
propose \centris (TLSH + SBB) as a variant of \centris (TLSH) with SBB
protected. It has two additional parameters for using the SBB protocol. $\gamma$
is the mutation bias for the client, and $\theta$ is the bucket size returned by
the server. In the experiment, we set $\gamma = 50$ and $\theta = 1,000$.

\parh{\centris (DPCNN).}~We enhance the native \centris by replacing TLSH with a
well-trained DPCNN model and improve the OSS identification process by assigning
weights to the source functions. To train the DPCNN model, we follow the
approach of Code2Vec~\cite{alon2019code2vec} by training the DPCNN model to
predict the name of a function with the function's implementation. All source
functions collected in our OSS database (see \S~\ref{subsec:oss-database}) are
split into training (80\%), validation (10\%), and testing (10\%) data.
We clarify that we do not choose widely used
CodeBert~\cite{feng2020codebert} and Code2Vec due to their inefficiency
and limited support of MPC frameworks.

\parh{\centris (DPCNN + \crypten).}~As introduced in \S~\ref{subsec:mpc-sca},
this variant uses \crypten to encrypt the client's tensors before sending them
to the server. It uses the same DPCNN model as \centris (DPCNN).
In the experiment, we set up two parties, the client and the server, and we use
the official interface of \crypten to encrypt the DPCNN model.

\parh{\binaryai}~\cite{binaryai2024,binaryaiurl} is a commercialized
binary-based SCA tool from the industry. We use its official service in an
out-of-the-box setting. \binaryai\ takes stripped binary executables
(without symbols/debug information) as its inputs,
which is reasonable as few closed-source binaries are unstripped.

\vspace{-2pt}
\subsection{Performance on SCA}
\label{subsec:performance-sca}

\begin{table}
    \vspace{-5pt}
    \caption{Precision (\textit{P}), recall (\textit{R}), F1 score (\textit{F1}), and time cost of various SCA tools.}
    \vspace{-5pt}
    \hspace{-10pt}
    \label{tab:accuracy1}
    \resizebox{1.05\linewidth}{!}{
    \begin{threeparttable}
    \begin{tabular}{c|c|ccc|cc}
        \hline
        Tool & Type$^{\text{1}}$ & \textit{P} & \textit{R} & \textit{F1} & Time (s) & Overhead \\ \hline
        \centris (TLSH)  & S   & .555 & .592 & .573 & 239.74 & - \\ \hline
        \centris   & S \&  & \multirow{2}{*}{.554} & \multirow{2}{*}{.593} & \multirow{2}{*}{.573} & \multirow{2}{*}{10149.45} & \multirow{2}{*}{4133\%} \\ 
        (TLSH + SBB) & SBB  &   &   &   &   &   \\ \hline
        \centris (DPCNN)  & S  & .838 & .769 & .802 & 715.21 & 198\% \\ \hline
        \centris  & S \& & \multirow{2}{*}{.838} & \multirow{2}{*}{.769} & \multirow{2}{*}{.802} & \multirow{2}{*}{131332.66} & \multirow{2}{*}{54681\%$^{\text{2}}$} \\ 
        (DPCNN + \crypten) & MPC &  &  &  &  &  \\ \hline
        
        \binaryai & B & .881 & .477 & .619 & NA$^{\text{3}}$ & NA \\
        \hline
        \multicolumn{7}{l}{$^{\text{1}}$ Tools are classified into source-based (S), binary-based (B), SBB-based (SBB), and} \\
        \multicolumn{7}{l}{\ \ MPC-based (MPC).} \\
        \multicolumn{7}{l}{$^{\text{2}}$ Compared with \centris (DPCNN), the overhead is 18283\%, nearly 184$\times$ slower.} \\
        \multicolumn{7}{l}{$^{\text{3}}$ We are unable to measure the time cost of \binaryai via its official web service.}
    \end{tabular}
    \end{threeparttable}
    }
    \vspace{-15pt}
\end{table}

\parh{Evaluation Metrics.}~Precision, recall, and F1 scores are used to evaluate
the accuracy of SCA tools. Those metrics are widely used in the SCA scenario~\cite{atvhunter,
centris,binaryai2024,libdb2022}. Moreover, since privacy protocols like
SBB and \crypten introduce additional overhead, we also measure the time cost
of each SCA tool.
In the experiment, we use the database constructed in \S~\ref{subsec:oss-database}
and the dataset introduced in \S~\ref{subsec:dataset} for evaluation.

\smallskip
\parh{Accuracy.}~\T~\ref{tab:accuracy1} presents the precision, recall, and F1
scores of \centris's variants and \binaryai. We observe that \centris (TLSH) and
\centris (TLSH + SBB) have similar accuracies. A similar phenomenon is observed
between \centris (DPCNN) and \centris (DPCNN + \crypten), denoting that privacy
protocols do not undermine the SCA analysis.
We also observed that \centris (DPCNN) largely outperforms \centris (TLSH).
We attribute the improvement to the weight enhancements in \centris (DPCNN)
elaborated in \S~\ref{subsec:centris-dpcnn}. Indeed, \centris (DPCNN) has a
similar accuracy to \centris (TLSH) when the weight enhancement is disabled.

By comparing the industrial binary-based SCA tool, \binaryai, with the source-based
\centris (DPCNN), we observe that \binaryai's precision is surprisingly high,
close to \centris (DPCNN). However, the recall is relatively low, leading to a
lower F1 score.
We noticed that \binaryai is much more accurate when analyzing binaries with
exported symbols,
and \binaryai's OSS database already contains the 14 OSS repositories
used as our test dataset. While this indicates certain ``unfairness,'' 
we are
unable to modify \binaryai as it is close-source.
However, as shown in \T~\ref{tab:accuracy1}, it still gains
relatively poor accuracy compared to analyzing
source code.

Overall, \centris (DPCNN) and \centris (DPCNN + \crypten) achieve the best
F1 score among all evaluated SCA tools, indicating that the source-based SCA
framework is accurate and effective, and the privacy-preserving protocols
introduced in \S~\ref{sec:pp-sca} do not notably affect the accuracy.

\smallskip
\parh{Time Cost.}~By comparing the time cost of \centris (TLSH) and
\centris (TLSH + SBB), we observe that the time cost is significantly
increased by $41\times$ due to the extra computations incurred by SBB.
Similarly, the time cost of \centris (DPCNN + \crypten) is
$184\times$ slower than \centris (DPCNN) due to the overhead of \crypten.
Thus, we conclude that the privacy-preserving protocols (i.e., SBB and \crypten)
introduce an enormous overhead to the SCA tools.

On the other hand, by comparing \centris (TLSH) and \centris (DPCNN),
we observe that the time cost of \centris (DPCNN) is nearly tripled due to the
embedding process. However, \centris (DPCNN) is still efficient enough for
large-scale SCA tasks, and more efficient hardware can further reduce
the overhead. We view the time cost of using DPCNN as
acceptable for SCA tasks.

\vspace{-2pt}
\subsection{Leakage of SBB-Based SCA}
\label{subsec:sbb-sca-leakage}

As mentioned in \S~\ref{subsec:sbb-sca}, although the client does not
send the original hash value $p$ to the server,
the mutated hash value $p'$ is still
close to the $p$. Thus, $p'$ may
disclose some information about $p$ to the server.
In this section,
we uncover the subtle privacy leakage of SBB-based SCA,
i.e., \centris (TLSH + SBB).
The experiment follows the semi-honest threat model, and we
assume the customer (holding the client) and vendor (holding the server)
follow the SBB protocol faithfully.
The vendor stores all mutated function hash values and
is curious about the client's SCA report.

Following the workflow of SBB-based SCA shown in \F~\ref{fig:sbb}, the client
first dissects the input (\ding{192}) and gets the hash value set $I$
(\ding{193}). During the SBB-based analysis, the client mutates the hash value
$h_i \in I$  with the mutation bias $\gamma$ to get $h_i'$, which is sent to the
server (\ding{194}). Thus, the server has access to the set of mutated hash
values $I'$, from which the curious vendor can predict the client's SCA report.

\begin{figure}[]
    \vspace{-5pt}
    \hspace{-10pt}
    \centering
    \begin{subfigure}[b]{0.49\linewidth}
        \centering
        \includegraphics[width=1.0\textwidth]{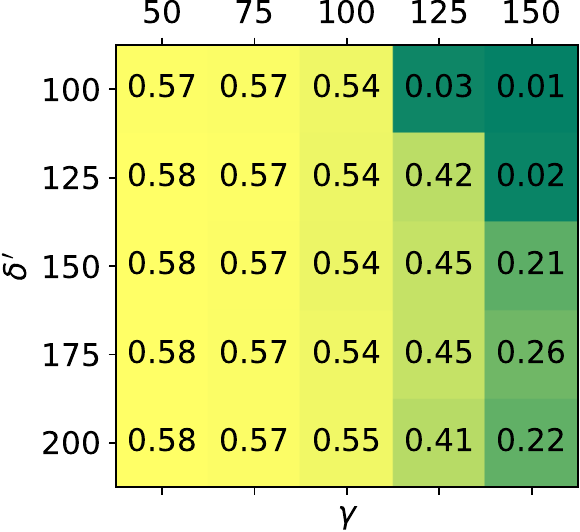}
        \vspace{-15pt}
        \caption{Client-side F1 scores.}
        \label{fig:sbb-client}
    \end{subfigure}
    \begin{subfigure}[b]{0.49\linewidth}
        \centering
        \includegraphics[width=1.0\textwidth]{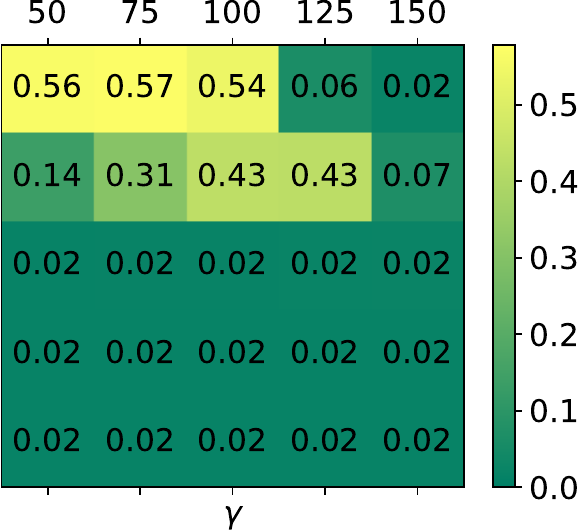}
        \vspace{-15pt}
        \caption{Server-side F1 scores.}
        \label{fig:sbb-server}
    \end{subfigure}
    \vspace{-5pt}
    \caption{SBB-based \centris's F1 according to $\gamma$ and $\delta'$.}
    \label{fig:sbb-leakage}
    \vspace{-15pt}
\end{figure}

Formally, let $DB_k$ be the hash value set of the $k$-th OSS project, and the
function $dis$ evaluates the distance between two hash values.
Given the mutated hash values set $I'$, the curious vendor can infer whether the
$k$-th OSS project is reused in the following way. Firstly, the vendor computes
a set $S_k'$ with $I'$, $DB_k$, and a threshold $\delta'$ as
\E~\ref{eq:centris2}.
\begin{equation}
    \hspace{-6pt} S_k' = \{(h_i', h_j) | \forall (h_i', h_j) \in I' \times DB_k\ \text{and}\ dis(h_i', h_j) < \delta' \}
    \hspace{-3pt} \label{eq:centris2}
\end{equation}
\noindent Secondly, by iterating all OSS projects and computing
$S_k' (k \in [1, n])$, the vendor can rank all OSS projects
by the size of $S_k'$, i.e., $|S_k'|$. A larger $|S_k'|$ denotes
a higher probability that the consumer's software reuses $k$-th OSS project.
Thus, the curious vendor can infer the client's privacy.
Intuitively, a larger mutation bias $\gamma$ increases the
uncertainty of the vendor's inference.
\F~\ref{fig:sbb-client} presents the relations between mutation bias ($\gamma$),
vendor's threshold ($\delta'$), and F1 scores of a client.
\F~\ref{fig:sbb-server} illustrates the predicted F1 scores of a server.
Visually, the yellow cells in the same positions of \F~\ref{fig:sbb-leakage}(a)
and \F~\ref{fig:sbb-leakage}(b) indicate
that the client's and server's F1 scores are close and high,
demonstrating the vendor can predict the SCA report,
thereby leaking the OSS projects reused by the client.

When the mutation bias $\gamma$ is relatively small,
the SCA service works functionally, and the client has decent F1 scores,
while a large $\gamma$ leads to a poor F1 score for the client.
To ensure the client's accuracy and privacy, the client's $\gamma$
should be small (i.e., yellow cells in \F~\ref{fig:sbb-client}),
and the vendor's threshold $\delta'$ should be large (i.e., green cells in
\F~\ref{fig:sbb-server}).
However, $\delta'$ is selected by the vendor.
From the vendor's perspective, it can intentionally use a relatively
small $\delta'$ (e.g., 100 or 125) for prediction. Thus,
the server-side F1 score is close to the client-side
F1 score, which means the vendor could infer the SCA results as well as the
client, posing a serious threat to the customer privacy protection of
SBB-based SCA.

Overall, by observing the different privacy guarantees and accuracies
between \centris (TLSH + SBB) and \centris (DPCNN + \crypten)
in \T~\ref{tab:accuracy1}, we believe that MPC-based SCA is
more suitable for privacy-preserving SCA.
Nevertheless, the MPC-based SCA incurs a high overhead, which
we will address in \S~\ref{sec:design}.

\vspace{-2pt}
\section{Optimizing MPC-Based SCA}
\label{sec:design}

\begin{figure*}
    \begin{minipage}[b]{0.55\linewidth}
        \centering
        \includegraphics[width=1.0\linewidth]{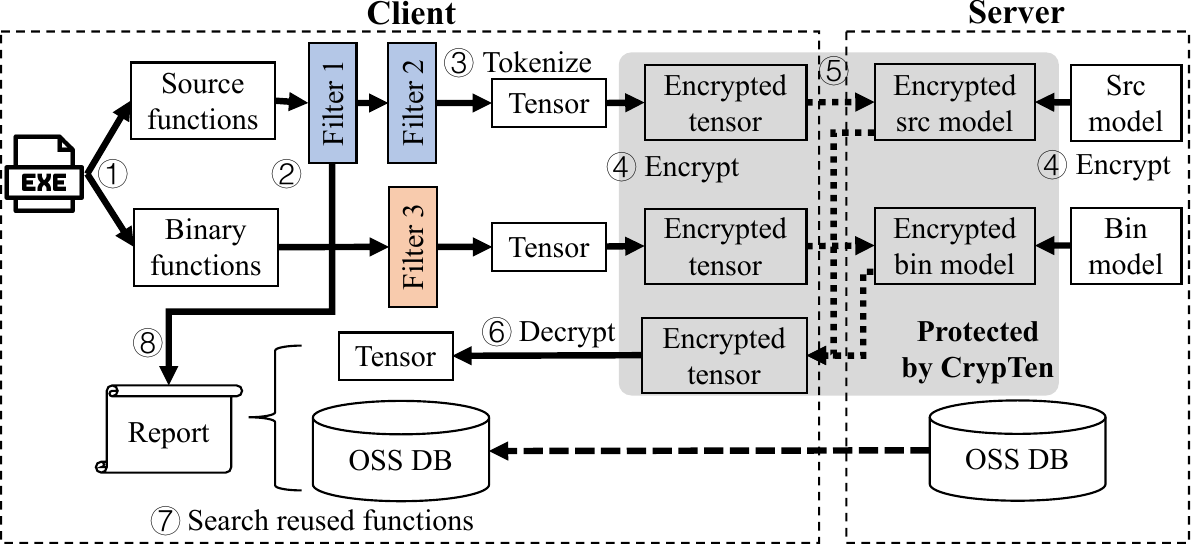}
        \vspace{-17pt}
        \caption{Workflow of \tool.}
        \label{fig:workflow}
    \end{minipage}
    \hfill
    \begin{minipage}[b]{0.35\linewidth}
        \centering
        \includegraphics[width=1.0\linewidth]{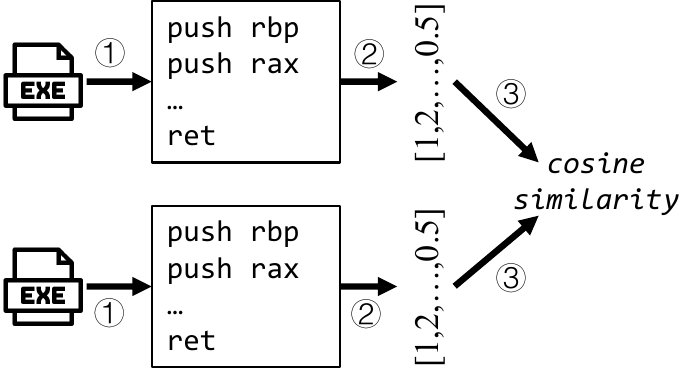}
        \vspace{-15pt}
        \caption{A common pipeline of binary code similarity analysis (BCSA).}
        \label{fig:bsa-pipeline}
    \end{minipage}
    \vspace{-20pt}
\end{figure*}

The promising result in \S~\ref{subsec:performance-sca} of the
MPC-based SCA framework in accuracy and principled privacy guarantee
motivates us to build a practical privacy-preserving SCA framework
based on MPC. However, the MPC-based SCA framework 
comes with a high overhead. This section optimizes the
standard setup of MPC-based SCA by reducing its cost.
The insight of this optimization is selecting only necessary data
with software analysis approaches for expensive encrypted computation;
thus, we significantly reduce the volume of data required for server queries
and achieve a reasonable overhead with high accuracy and privacy guarantee.

\smallskip
\parh{Assumption.}~We assume the
customers of \tool\ are developers, and the client has access to
the \textit{source code} and the compiled \textit{binary executables} with debug
information.
The assumption is reasonable due to the following reasons:
(1) Since the MPC-based SCA does not disclose any clients'
information, utilizing both the source and binary code does not harm clients' privacy.
(2) Consequently, \tool can access the developing environment safely;
hence, the developers can easily provide \tool with the source code and the
building products (i.e., binary executables).
(3) Building products with debug information for testing is a common
practice~\cite{zhou2024plankton}, and symbols are usually available. 
As a result, by analyzing both the source and
binary code, \tool can leverage the source code to achieve better accuracy and
identify the reused components without source code, e.g., statically linked
libraries.

\smallskip
\parh{Workflow.}~\F~\ref{fig:workflow} presents the workflow of
\tool. Like \centris (DPCNN + \crypten), a client needs to download the OSS DB first.
Given the input binary with debug information, we
preprocess it to extract functions from the source codebase, i.e., ``source functions'',
and functions reused without source code, i.e., ``binary functions'' (\ding{192}).
Then, the functions are filtered and
selected for expensive encrypted computation (\ding{193}). The selected source
functions are processed by the source embedding model, which is the same as the
process of \centris (DPCNN + \crypten). Similarly, the selected binary functions
are processed by the binary embedding model (\ding{194}-\ding{197}). To generate
the SCA report, the client searches for the reused functions in the database
(\ding{198}) and combines the reused OSS knowledge detected by the symbol filter
(\ding{199}).

Compared with the workflow of \centris (DPCNN + \crypten)
(\F~\ref{fig:mpc}), the main optimizations are conducted on the client side.
\tool's workflow is more complex and involves an extra binary embedding model
(\S~\ref{subsec:binary-embedding}) and three filters
(\S~\ref{subsec:client-filters}). Besides, we also accelerate the offline
signature generation process, as will be discussed in
\S~\ref{subsec:improve-sig}.

\vspace{-2pt}
\subsection{Binary Embedding Model}
\label{subsec:binary-embedding}

Similar to embedding source code for analysis, we need to embed binary code pieces into embedding vectors.
Fortunately, many existing binary code similarity analysis techniques (BCSA) can
produce function-level embeddings to capture the semantics of
assembly code~\cite{yu2020order,yu2020codecmr,massarelli2019safe,ding2019asm2vec,
sem2vec2023,binuse2022,binaug2024,li2022unleashing}.
Therefore, we can leverage BCSA in \tool's workflow.
\F~\ref{fig:bsa-pipeline} presents a frequently used pipeline of binary
similarity analysis frameworks. Given two binary executables, a BCSA tool
disassembles binaries to assembly code and produces the embedding
vectors of assembly functions. The similarity score is usually computed as the
cosine similarity of embeddings.

Following the training pipeline presented by recent research in relevant topics,
\gemini~\cite{xu2017neural} and \safe~\cite{massarelli2019safe}, we train a
DPCNN model to embed binary code. We report that the AUC score, a widely-used
metric for binary similarity
works~\cite{massarelli2019safe,li2021palmtree,yu2020codecmr,yu2020order}, of our
binary code embedding model is 0.935, which is comparable to \gemini (0.932) and
\safe (0.990), demonstrating the effectiveness of our setting.
We, however, do not use an existing model due to implementation difficulties.
Existing binary code embedding techniques often utilize graph
structures~\cite{yu2020order,
ding2019asm2vec,ncc,duan2020deep,zuo2019neural,xu2017neural,wang2022jtrans} and
recurrent neural networks~\cite{massarelli2019safe} to embed assembly code.
However, we find that few (if at all any) existing crypto protocols support
those models due to computation complexity.

\vspace{-2pt}
\subsection{Improvements in Signature Generation}
\label{subsec:improve-sig}
Existing works (e.g., \centris and \tplite) treat each \texttt{tag} of a Git
repository as a unique version. Given a repository's specific version, they
analyze its code and generate signatures with extracted features (e.g., strings,
magic numbers, functions, and file hashes) for the given version.
However, since real-world OSS developers may not treat a tag as a release version,
their approach is impractical for our complex OSS database due to the huge cost.
For instance, \texttt{ClickHouse}~\cite{clickhouseurl} generates a new tag every
week, resulting in thousands of versions; consequently, generating signatures for
\texttt{ClickHouse} can take more than four months, per our estimation.

As a practical setting, we build signatures of a repository by commits. Given a
repository, after generating the signature for the first version, we focus on
the files changed by the subsequent commits, whereas the unchanged files will
not be analyzed again. This way, the OSS signature generation process is much
more efficient.
Following existing works (\centris and \ossfp~\cite{wu2023ossfp}), we extract
function-level features. Thus, the changes between different
commits will include newly created functions, changed functions, and deleted
functions. In this study, a function's identifier (i.e., ID) is composed of its
name and scope; hence, a newly created ID denotes a new function. Functions with
the same ID but different implementations between two commits are changed
functions. A missing ID denotes a deleted function. Moreover, we also record the
direct call relations between two functions, which results in a large call graph
for each repository.

\vspace{-2pt}
\subsection{Client-Side Filters}
\label{subsec:client-filters}

As shown in \F~\ref{fig:workflow}, after preprocessing an input binary and
extracting its source and binary functions, the client will tokenize those
functions for expensive encrypted computation. To reduce the overhead incurred
by crypto, we propose three filters to identify suspect functions on the
client side before tokenization and analysis.
Specifically, the symbol filter first conservatively pinpoints functions copied
from OSS projects; such identified functions do not go through subsequent SCA
analysis. For the remaining functions that are not identified, the source
function filter and assembly function filter are employed to pick representative
functions to be tokenized and embedded for further SCA analysis.

\smallskip
\parh{Filter 1: Symbol Filter.}~\tool's client can safely
employ sensitive information, including
symbols since all data is encrypted before being sent to the server.
Therefore, given the symbols provided by customers, a
straightforward idea is to pinpoint functions that are likely copied from
existing OSS projects by matching their symbols, such as function names and
scopes. To do so, we also collect names and scopes (usually namespace names) of
functions while building the OSS database. Ideally, given an input function, if
it has identical symbols with one record in the OSS database, it is likely to be
a reused function, and we forward it to further analysis.

However, directly collecting accurate symbol names (including scope names) from
the source code is difficult due to the complexity of C/C++ projects.
Strictly matching symbols (i.e., names and scopes) will likely lead to an
extensive amount of false negatives (FNs). Thus, we add a more flexible rule as
a remedy for symbol matching to reduce FNs. Given that matches of complex (e.g.,
over 20 characters) and unique function names are usually strong indicators of
direct function reuses, we view a function as copied from OSS projects if it has
a complex name and is matched to record in the database. Besides, to increase
the accuracy of name matching, we further check the function call relationship
between potentially matched functions. A function is identified as reused from
an OSS project only if the function and its matched function have the same
call relationship, i.e., both have the same edge in the call
graph.

Our evaluation shows that the proposed symbol filtering method can reach an F1
score of 0.547. Some reused functions cannot be recognized with symbol matching
because of modifications in function names and implementations during code
reusing. In order to locate the reused components that are not matched in this
step, we further involve source- and binary-based embedding models to analyze
the remaining functions. 

\smallskip
\parh{Filter 2: Informative Source Function Filter.}~When the source code is available,
we tokenize and embed source functions for further similarity-based SCA. Before
that, the informative source filter is used to determine and skip
inconsequential functions. One insight behind this filter is that some trivial
functions hardly contribute to the SCA results. Thus, we can only consider
complex and representative functions for time-consuming analysis.

Given functions that are not matched by the symbol filter, we select those that
are informative and embed their source code for SCA analysis. Specifically, we
use the maintainability index~\cite{maintainabilityindexurl,wu2023ossfp} to
decide if a function is ``informative.'' In short, this metric is negatively
correlated to the cyclomatic complexity~\cite{shepperd1988critique}, Halstead
volume~\cite{hariprasad2017software},
and the lines of code in a function. We configure our tool to keep a
proportion of functions of each source file with the least maintainability indexes
for further analysis. For simplicity, we define the proportion of kept
functions as the hyperparameter $\theta_1$.

\smallskip
\parh{Filter 3: Assembly Function Filter.}
Similar to Filter 2, when a binary is provided for SCA, we first
disassemble it into assembly code and then identify representative functions for
embedding and analysis. However, binary code analysis is less accurate than
source code analysis in general. Therefore, the assembly function filter more
strictly restricts the conditions for picking a function.
It selects functions whose names exist in the collected OSS projects.
Also, we avoid selecting overly long functions due to the limitation of existing
binary code similarity analysis techniques on lengthy
input~\cite{liu2018diff,jia20231,massarelli2019safe,marcelli2022machine}.
In particular, to evaluate if an assembly
function should be selected, we use the following equation, 
\begin{equation}
    \hspace{-2pt} score = {1} / {(ln(OSS\_num + 2) * ln(block\_num + 2))}
    \hspace{-3pt} \label{eq:bin-select}
\end{equation}
where $OSS\_num$ is the number of OSS projects containing the function's name.
$block\_num$ is the number of basic blocks in the function. We add 2 to the
denominator to avoid zero division. We also define another hyperparameter
$\theta_2$ to decide the proportion of functions for embedding and SCA analysis.

\vspace{-3pt}
\subsection{Privacy Analysis of \tool}
\label{subsec:privacy-tool}
Similar to \centris (DPCNN + \crypten), \tool\ protects the client's privacy
with the MPC protocol. The client's valuable assets (e.g., source and binary
code) are not leaked to the server.
From the server's perspective, the client downloads a symbol database to enable
the symbol filter (Filter 1), and the OSS DB also contains the binary embeddings of OSS
projects. However, the symbol database only contains function names and
namespaces. We thus argue that the symbol database is not a highly valuable
asset since it can be easily obtained by parsing an OSS project. Additionally,
merely using the symbol database cannot achieve a high-quality SCA result
(i.e., 0.547 F1 score as evaluated in \S~\ref{subsec:impacts-filters}).
Regarding the OSS DB, the binary embeddings are used the same way as the source
embeddings; therefore, the OSS DB does not leak any additional information since
the client has no access to the code embedding model.



\vspace{-2pt}
\section{Evaluation}
\label{sec:evaluation}

To evaluate \tool, we reuse our large-scale OSS database described
in \S~\ref{subsec:oss-database} and the manually-built evaluation  
dataset (\S~\ref{subsec:dataset}).
We also reuse the source-based OSS scoring method and
source code embedding model of \centris (DPCNN)
(see \S~\ref{subsec:performance-sca}) for \tool.
\S~\ref{subsec:self-performance} shows the impact of threshold value $\beta$ on
\tool's performance. To ensure the efficiency of \tool,
we use a small threshold $\theta_1 = 0.02$ for the informative source function
filter, and $\theta_2 = 0.02$ for the assembly function filter.
Due to the space limit, we present a detailed analysis of the impact of
$\theta_1$ and $\theta_2$ on \tool's performance on our website~\cite{snapshot}.

\vspace{-3pt}
\subsection{Performance of \tool}
\label{subsec:self-performance}

\begin{figure}
    \vspace{-10pt}
    \centering
    \begin{minipage}{0.480\linewidth}
        \centering
        \includegraphics[width=1.03\textwidth]{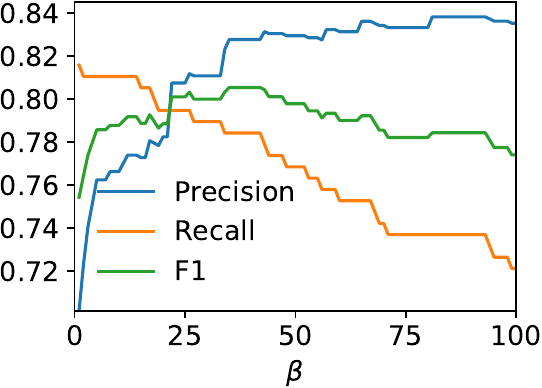}
        \vspace{-15pt}
        \caption{Precision, recall, and F1 score of \tool with different
        values of $\beta$.}
        \label{fig:perf}
    \end{minipage}
    \hfill
    \begin{minipage}{0.480\linewidth}
        \centering
        \includegraphics[width=1.03\textwidth]{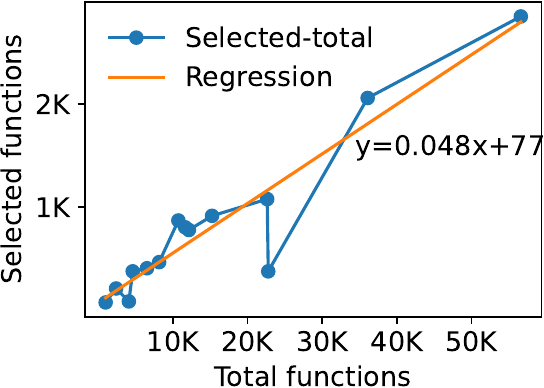}
        \vspace{-15pt}
        \caption{Number of selected functions of \tool to
        total functions of each binary.}
        \label{fig:cost}
    \end{minipage}
    \vspace{-15pt}
\end{figure}

\F~\ref{fig:perf} presents the precision, recall, and F1 score of \tool with
different $\beta$ values. \tool's performance is not sensitive
to $\beta$'s changes, especially the F1 score (i.e., \textcolor{teal}{{green}}
line). When $\beta$ is greater than 5, \tool's F1 score is greater than 0.78.
The highest F1 score is 0.81 when $\beta$ is 40. Compared with \centris (DPCNN),
\tool achieves a similar F1 score, which demonstrates the effectiveness
of our proposed three filters.
Comparing \tool with and without optimizations (see \T~\ref{tab:accuracy2}),
we observe that the proposed optimizations do not sacrifice accuracy.

\F~\ref{fig:cost} presents the total number of input functions and the number of
functions \tool\ selected to perform encrypted computation. The regression line
(i.e., \textcolor{orange}{orange} line) has a 0.048 gradient, indicating that
\tool selects less than 5\% of functions to generate embeddings for SCA.
As shown in \T~\ref{tab:accuracy2},
compared with \tool without optimization (i.e., all functions are
passed to the embedding models), three filters reduce the time
cost to 8.47\%, i.e., 11.8$\times$ faster. Compared with naively using
MPC protocol in the SCA framework (184$\times$ overhead),
\tool's overhead is merely 23$\times$.
In total, \tool takes 4.6 hours to identify reused OSS projects of our dataset
(i.e., 20 minutes per binary on average), demonstrating the scalability of
\tool. Our dataset's largest binary,
\texttt{mod_pagespeed}~\cite{modpagespeedurl}, whose repository reaches 543 MB
without history files, takes 72 minutes for \tool.

\vspace{-4pt}
\subsection{Impacts of Filters}
\label{subsec:impacts-filters}

As we design three filters to accelerate \tool and
report the effects of enabling all of them,
we further analyze the impact of each filter separately.
The overhead of each filter is negligible compared with
computing the embedding vectors; thus, we focus on analyzing the accuracy
impacts of each filter.

\smallskip
\parh{Filter 1.}~After extracting functions and their symbols from an input
software, we first analyze the software with the symbol filter, which is
expected to have high precision since we only consider representative symbols
and require their direct function calls to be matched. In other words, the symbol
filter is conservative and only rules out functions that can be skipped with
high confidence. Our experiments show that applying the symbol filter alone
(i.e., \ding{192}\ding{193}\ding{199} of \F~\ref{fig:workflow} only;
no functions are analyzed by {Filter 2} and {Filter 3})
achieves 0.918 precision but 0.4 recall, resulting in a 0.547 F1 score.
Below, we show that the following two filters can further improve the SCA
accuracy based on the results of the symbol filter.

\smallskip
\parh{Filter 2.}~After {Filter 1},
the remaining source functions are sent to {Filter 2} for further
analysis with the source embedding model. To evaluate the effect of Filter 2,
{Filter 3} and the following binary embedding model are not used.
We report that SCA's precision and recall are 0.829 and
0.747 when $\beta=40$. Similar to the conclusion in
\S~\ref{subsec:self-performance}, changing $\beta$ has little impact on the F1
score at this step.
This filter can effectively improve the recall (from 0.40 to 0.747),
achieving similar accuracy to analyzing all source functions (i.e., \centris (DPCNN)).

\smallskip
\parh{Filter 3.}~This filter applies to disassembled assembly
functions when binary is provided as input. Functions selected by this filter
will be embedded into vectors using BCSA techniques.
A function is identified as reused from an OSS project when its matched symbol
and the most similar embedding vector are both from that OSS repository. This
requirement facilitates \tool\ in eliminating false positives caused by assembly
function embeddings. Our evaluation reveals that it improves the recall from
0.747 to 0.791. While the improvement is limited due to the difficulty of
analyzing binary code, we view it as a critical replenishment when binary is the
only input.

\begin{table}[]
    \center
    \vspace{-10pt}
    \caption{Performance of \tool\ with different settings.
    }
    \label{tab:accuracy2}
    \vspace{-5pt}
    \resizebox{1.0\linewidth}{!}{
    \begin{threeparttable}
    \begin{tabular}{c|ccc|c|c}
    \hline
     & \textit{P} & \textit{R} & \textit{F1} & Time cost (s) & Cost ratio \\\hline
     \tool (w/o opt)$^{\text{1}}$ & .802 & .780 & .791 & 195125.06 & 100\% \\
     \hline
     \tool\ & .828 & .791 & .809 & 16527.3 & 8.47\% \\  \hline
     \centris (DPCNN)$^{\text{2}}$ & .838 & .769 & .802 & 715.2 & 0.37\% \\ 
     \hline
    \end{tabular}
    \begin{tablenotes}
        \item[1] \tool\ (w/o opt) denotes \tool without three filters.
        \item[2] To ease comparison with \T~\ref{tab:accuracy1},
        we present \centris (DPCNN).
    \end{tablenotes}
    \end{threeparttable}
    }
    \vspace{-15pt}
\end{table}

\subsection{False Positives (FPs) and False Negatives (FNs)}
\label{subsec:analysis-fp-fn}

Static security analyzers can hardly avoid FPs and
FNs~\cite{codeql,joern,bdbaurl,binaryaiurl,DBLP:conf/icse/WuHHWTS0024,
DBLP:journals/tosem/WangWYSZXZ23}. Nevertheless, as shown in
\T~\ref{tab:accuracy2}, \tool\ has already achieved
the \textit{best F1 score} with acceptable overhead compared with existing
SCA solutions (without privacy-preserving).
\tool's embedding-based analysis is the major root cause of FPs and FNs.
To reduce them, we advocate for the privacy community to better support more
advanced embedding techniques (e.g., more powerful models).
Also, our manual analysis finds that the imperfect OSS database also
contributes to FPs and FNs.

\parh{Missed OSS Projects.}~Although we have collected over 1,800 OSS projects by
going through the references of CVEs, our OSS database still misses some OSS
projects. For instance, \texttt{libunwind}~\cite{libunwindurl} is not included
in our OSS experiment, but it is reused by \texttt{crashpad}~\cite{crashpadurl}.
When we perform SCA on \texttt{mod\_pagespeed}~\cite{modpagespeedurl},
\texttt{crashpad} is an FP since
\texttt{mod\_pagespeed}
shares \texttt{libunwind}'s functions with \texttt{crashpad}. To form the
current OSS database in this research, we have already spent a significant
amount of time and resources (over 800 GB of OSS repositories);
industrial audiences of this paper may
consider putting more effort into augmenting the OSS database and
reducing FPs and FNs.

\parh{Generative Code Reuse.}~Following previous studies (e.g., \centris and
\tplite), this work considers four types of OSS reuses, i.e., exact reuse,
partial reuse, structural-changed reuse, and code-changed reuse~\cite{centris}.
While these four categories cover most cases, this study finds a
new rare type of reuse, i.e., generative code reuse. This type of reused code is
generated dynamically by a script; thus, the generated code is not analyzed
during OSS signature generation. For instance, \texttt{yuzu}~\cite{yuzuurl}
and \texttt{raylib}~\cite{rayliburl}
reuse code generated by \texttt{glad}~\cite{gladurl}. Therefore, when analyzing programs
compiled from \texttt{yuzu}, the code borrowed from \texttt{glad} in \texttt{yuzu}
is identified as part of code from \texttt{raylib}, resulting in an FN (i.e.,
missing \texttt{glad}) and an FP (i.e., \texttt{raylib} is identified as a
component of \texttt{yuzu}).

\vspace{-2pt}
\section{Related Work}
\label{sec:related}

Recent SCA works target different programming languages and different types of
software. However, these tools do not naturally offer a privacy-preserving
solution.
Conventional binary-based SCA tools~\cite{bat2011,osspolice,tplite2023issta}
primarily rely on string-level signatures. \bsfinder~\cite{b2sfinder}
additionally uses \texttt{if/else} and \texttt{switch} structures. However,
these signatures may not exist in some OSS projects (e.g.,
RapidJson~\cite{rapidjsonurl}) and are not robust to code changes. Recent
binary-based SCA~\cite{binaryaiurl,libdb2022,yang2022modx} also explores using
embedding techniques, which are often limited by the hurdles of reverse
engineering.
Source-based SCA tools~\cite{centris,wu2023ossfp,sourcerercc} for C/C++
software usually rely on code-block- and function-level signatures.
\cite{sourcerercc} uses code-block-level signatures and
is relatively time-consuming. Recent works use function-level signatures.
\centris~\cite{centris} uses code segmentation to reduce FPs and leverages
redundancy elimination to reduce the size of the OSS database.
\ossfp~\cite{wu2023ossfp} employs the maintainability index to reduce the OSS
database's size and improve SCA scalability. Our work also uses the
maintainability index to reduce the expensive encrypted computation.

\vspace{-2pt}
\section{Conclusion}
\label{sec:conclusion}

We have presented the first study on privacy-preserving SCA. We review SCA
privacy requirements and depict viable technical solutions with varying privacy
gains and overheads. We also empirically benchmark the proposed solutions and
discuss issues in privacy leakage and cost. Accordingly, we optimize the
MPC-based SCA by reducing its overhead significantly without
sacrificing the privacy guarantee or accuracy. This work provides guides for
researchers and users who aim to use and improve SCA in practice.


\vspace{-2pt}
\section{Acknowledgments}
\label{sec:acknowledgments}
This work was supported in part by CCF-Tencent Open Research
Fund. We are grateful to the anonymous reviewers for their
valuable comments.

\bibliographystyle{IEEEtranS}
\bibliography{bib/machine-learning,bib/decompiler,
bib/timing,bib/sidechannel,bib/analysis,bib/ref,
bib/bsgx,bib/testing-cv,bib/cv,bib/sca,
bib/similarity,bib/privacy,bib/llm,bib/model-stealing}

\end{document}